\documentstyle[epsfig]{mn}
\newcommand{\etal}{{et~al.}}

\newcommand{\ro}{{\sl ROSAT}}

\newcommand{\eu}{{\sl EUVE}}
\newcommand{\iu}{{\sl IUE}}
\newcommand{\hs}{{\sl HST}}
\newcommand{\orf}{{\sl ORFEUS}}

\newcommand{\hei}{\mbox{$\rm {He\,{\sc i}}\:$}}
\newcommand{\heii}{\mbox{$\rm {He\,{\sc ii}}\:$}}

\newcommand{\ciii}{\mbox{$\rm {C\,{\sc iii}}\:$}}

\newcommand{\ci}{\mbox{$\rm {C\,{\sc i}}\:$}}
\newcommand{\civ}{\mbox{$\rm {C\,{\sc iv}}\:$}}
\newcommand{\ov}{\mbox{$\rm {O\,{\sc v}}\:$}}
\newcommand{\oiv}{\mbox{$\rm {O\,{\sc iv}}\:$}}
\newcommand{\nv}{\mbox{$\rm {N\,{\sc v}}\:$}}
\newcommand{\siii}{\mbox{$\rm {Si\,{\sc ii}}\:$}}
\newcommand{\siiv}{\mbox{$\rm {Si\,{\sc iv}}\:$}}
\newcommand{\svi}{\mbox{$\rm {S\,{\sc vi}}\:$}}
\newcommand{\fev}{\mbox{$\rm {Fe\,{\sc v}}\:$}}

\newcommand{\niv}{\mbox{$\rm {Ni\,{\sc v}}\:$}}

\newcommand{\tlus}{\mbox{$\rm {\sc tlusty}\:$}}
\newcommand{\pro}{\mbox{$\rm {\sc pro2}\:$}}
\newcommand{\iraf}{\mbox{$\rm {\sc iraf}\:$}}

\newcommand{\syn}{\mbox{$\rm {\sc synspec}\:$}}
\newcommand{\xsp}{\mbox{$\rm {\sc xspec}\:$}}

\title[Photospheric nickel in the hot DO white dwarf
REJ0503$-$289]
{The discovery of photospheric nickel in the hot DO white dwarf
REJ0503$-$289\thanks{Based on observations made with the Goddard High 
Resolution Spectrograph
on board the Hubble Space Telescope}}

\author[M.A. Barstow et al.]{
M.A. Barstow$^{1}$, S. Dreizler$^{2}$, J.B. Holberg$^3$, D.S.
Finley$^4$, K. Werner$^{2}$, I. Hubeny$^{5}$
\newauthor and E.M. Sion$^6$
\\
$^1$ {\it Department of Physics and Astronomy, University of Leicester,
University Road, Leicester LE1 7RH, UK}\\
$^2$ {\it Institut f\"ur Astronomie und Astrophysik, Universit\"at
Tubingen, Waldh\"auser Strasse 64, D-72076, T\"ubingen, Germany}\\
$^3$ {\it Lunar and Planetary Laboratory, University of Arizona, Tucson, 
AZ 85721, USA}\\
$^4$ {\it Eureka Scientific Inc., 2452 Delmer St., Suite 100 Oakland, CA
94602 }\\
$^5$ {\it Laboratory for Astronomy and Solar Physics, NASA/GSFC, Greenbelt,
Maryland, MD 20711, USA}\\
$^6$ {\it Department of Astronomy and Astrophysics, Villanova University,
Villanova, PA 19085, USA}
}

\begin{document}

\label{firstpage}

\maketitle

\begin{abstract}

We present the first evidence for the direct detection of nickel in the
photosphere of the hot DO white dwarf REJ0503$-$289. While this element
has been seen previously in the atmospheres of hot H-rich white dwarfs,
this is one of the first similar discoveries in a He-rich object. Intriguingly,
iron, which is observed to be more abundant than Ni in the hot DA stars, is not
detected, the upper limit to its abundance (Fe/He$=10^{-6}$) implying
a Fe/Ni ratio a factor 10 lower than seen in the H-rich objects
(Ni/He$=10^{-5}$ for REJ0503$-$289). 
The abundance of nickel and various other elements heavier than He were
determined from GHRS spectra. We used two completely independent sets of
NLTE model atmospheres which both provide the same results. This not
only reduces the possibility of systematic errors in our analysis but is also
an important consistency check for both model atmosphere codes.

We have also developed a more
objective method of determining $T_{\rm eff}$ and log g, from the He lines
in the optical spectrum, in the form of a formal fitting of the line
profiles to a grid of model spectra, an analogue of the
standard procedure utilising the Balmer lines in DA white dwarfs.
This gives the assigned uncertainties in $T_{\rm eff}$ and log g a firm
statistical basis and allows us to demonstrate that inclusion
of elements heavier than H, He and C in the spectral calculations, 
exclusively considered in most published optical analyses, yields 
a systematic downward shift in the measured value of $T_{\rm eff}$.

\end{abstract}


\section{Introduction}

About one quarter of all white dwarfs have helium-rich photospheres and
are classified according to the relative strengths of \heii \ and \hei \ 
lines in their optical spectra. These are determined by the ionization
balance of the He plasma, depending on the effective temperature of the
star. The DB white dwarfs display only \hei \ absorption lines, and cover
the temperature range from $\approx 11000-30000$K. Temperatures below
11000K are too low to excite \hei \ sufficiently to yield observable
lines, leading to the featureless DC white dwarfs.  In contrast, the hot DO
stars contain \heii \ lines alone, the ionization of helium
requiring higher effective temperatures than found in the DB white
dwarfs. The upper limit to the DO temperature range, at approximately
120000K, is associated with the helium, carbon and oxygen-rich PG1159 stars
(also denoted as DOZ by Wesemael \etal \ 1985, WGL), which are the
proposed precursors of the DO white dwarfs. The 45000K lower temperature
limit of the DO range is some 15000K higher than the beginning of the DB
sequence, presenting a so-called DO-DB gap between 30000K and 45000K,
first noted by Liebert \etal \ (1986). Subsequent surveys of white dwarfs
have failed to find examples of He-rich objects within the gap which continues to
present a problem in our understanding of white dwarf evolution.


It is clear, from the existence of the DO-DB gap and the changing ratio
of H-rich to He-rich objects, that white dwarf 
photospheric compositions evolve as the
stars cool. A number of physical mechanisms may be operating. For
example, the presence of elements heavier than H or He in white dwarf
photospheres is the result of radiative forces acting against the
downward pull of gravity, preventing these heavy elements sinking out of
the atmosphere (e.g. Chayer, Fontaine \& \ Wesemael 1995).  A possible
explanation of the DO-DB gap is that, following the AGB and PN phases of
mass-loss, residual hydrogen mixed in the stellar envelope floats to the
surface converting the DO stars into DAs. Later, the onset of convection
may mix the hydrogen layer, depending on its thickness,
back into the He-rich lower layers, causing the stars to
reappear on the DB sequence.

Any understanding of the possible evolutionary processes depends on
several important measurements, including the determination of effective
temperature, surface gravity and photospheric composition. Several
detailed studies have been carried out for DA white dwarfs (e.g. Marsh
\etal \ 1997; Wolff \etal
\ 1998;  Holberg \etal \ 1993, 1994; Werner \& \ Dreizler 1994),
but comparatively little work has been carried out on the DO stars. This
is partly due to the smaller number of stars available for detailed
study, but also arises from the comparative difficulty of establishing a
reliable self-consistent temperature determination from the \heii \
lines. The most recent and probably the most detailed study of the DO
white dwarf sample has been carried out by Dreizler and Werner (1996).
They applied the results of non-LTE model atmosphere calculations to the
available optical and UV spectra to determine the atmospheric parameters
of 14 stars, confirming the existence of the DO-DB gap. Dreizler and
Werner found the mean mass of the white dwarfs in their sample to be
$0.59\pm 0.08\rm M_\odot$, very close to the mean masses of the DA and DB
samples. A large scatter in heavy element abundances was found, even for
stars with similar parameters, and no clear trend along the cooling
sequence could be seen.

%
%
%

With such a small sample of DO white dwarfs available for study, each
individual object is significant. Discovered as a result of the \ro \ WFC
all-sky survey in the EUV, the DO white dwarf REJ0503$- $289 
(WD0501$-$289, MCT0501$-$2858) is
particularly important because of its low interstellar column density
(Barstow \etal \ 1994). As a consequence, it is the only DO white dwarf
which can be observed throughout the complete spectral range from optical
to X-ray and has been the subject of intense study. However, despite this
attention, it has not proved possible to generate a model spectrum that
is consistent at all wavelengths. Optical determinations of the effective
temperature, from a single ESO NTT spectrum yield a value of $\approx
70000$K, with a log surface gravity of 7.5 (Barstow \etal \ 1994;
Dreizler \& \ Werner 1996). Similar values are obtained from an LTE
analysis of the far-UV \heii \ lines in an \orf \ spectrum of the star
(Vennes \etal \ 1998).  In contrast, a lower temperature of 63000K is
required to reproduce the flux level of \eu \ spectrum and simultaneously
match the \ciii \  and \civ \  line strengths in the \iu \ high dispersion
spectra (Barstow \etal \ 1996). However, such a low temperature is
incompatible with the absence of \hei \ 4471\AA \ 
and \hei \ 5876\AA \ absorption lines in
the optical spectrum, which provides a lower limit on $T_{\rm eff}$ of
$\approx 65000$K.

Analyses of \iu \ high dispersion spectra yield measurements of the
abundances of C (also obtained from the optical spectrum), N, O and Si
and give limits on the presence of Ni and Fe. More recently, phosphorus
has also been detected in the REJ0503$-$289 \orf \ spectrum (Vennes \etal
\ 1998). The presence of heavy elements in the atmosphere of any white
dwarf, DO or DA, is known to have a significant effect on the temperature
structure of the photosphere and the emergent spectrum. With millions of
absorption lines in the EUV wavelength range, the influence of iron and
nickel is particularly dramatic. This is illustrated very clearly in the
DA stars, where the Fe and Ni opacity produces a steep drop in the
observed flux, compared to that expected from a pure H atmosphere (e.g.
Dupuis \etal \ 1995; Lanz \etal \ 1996; Wolff \etal \ 1998).  In
addition, for an H-rich model atmosphere including significant quantities of Fe
and Ni, the change in atmospheric structure also alters the predicted
Balmer line profiles. 
Inclusion of these effects, together with a NLTE analysis, has the
affect of yielding lower Balmer line effective temperatures compared
with those determined from a pure H-model atmosphere.  This results in
a net downward shift of the temperature scale for the hottest
heavy element-rich
objects (Barstow Hubeny \& \ Holberg 1998).

Compared to the extensive studies of the effect of Fe and Ni (and other
elements) on the atmospheres of DA white dwarfs, as discussed above,
little has been done in the case of the DO stars. First, there are few
detections of these species in the atmospheres of the DOs (see Dreizler
\& \ Werner 1996). Second, it has been more difficult to calculate
suitable stellar model atmospheres for comparison with the data. However,
it is possible that their inclusion in such computations might eventually
solve the probem of the EUV flux. We present an analysis of \hs \
spectra of REJ0503$-$289 obtained with the GHRS, which reveal the
presence of Ni in the atmosphere of the star, but yield only upper limits
to the abundance of Fe. We analyse a recent optical spectrum of the star,
to determine the effective temperature and surface gravity, and evaluate
the possible influence of photospheric Ni and trace Fe on the estimated
temperature.

\begin{figure*}
\leavevmode\epsffile{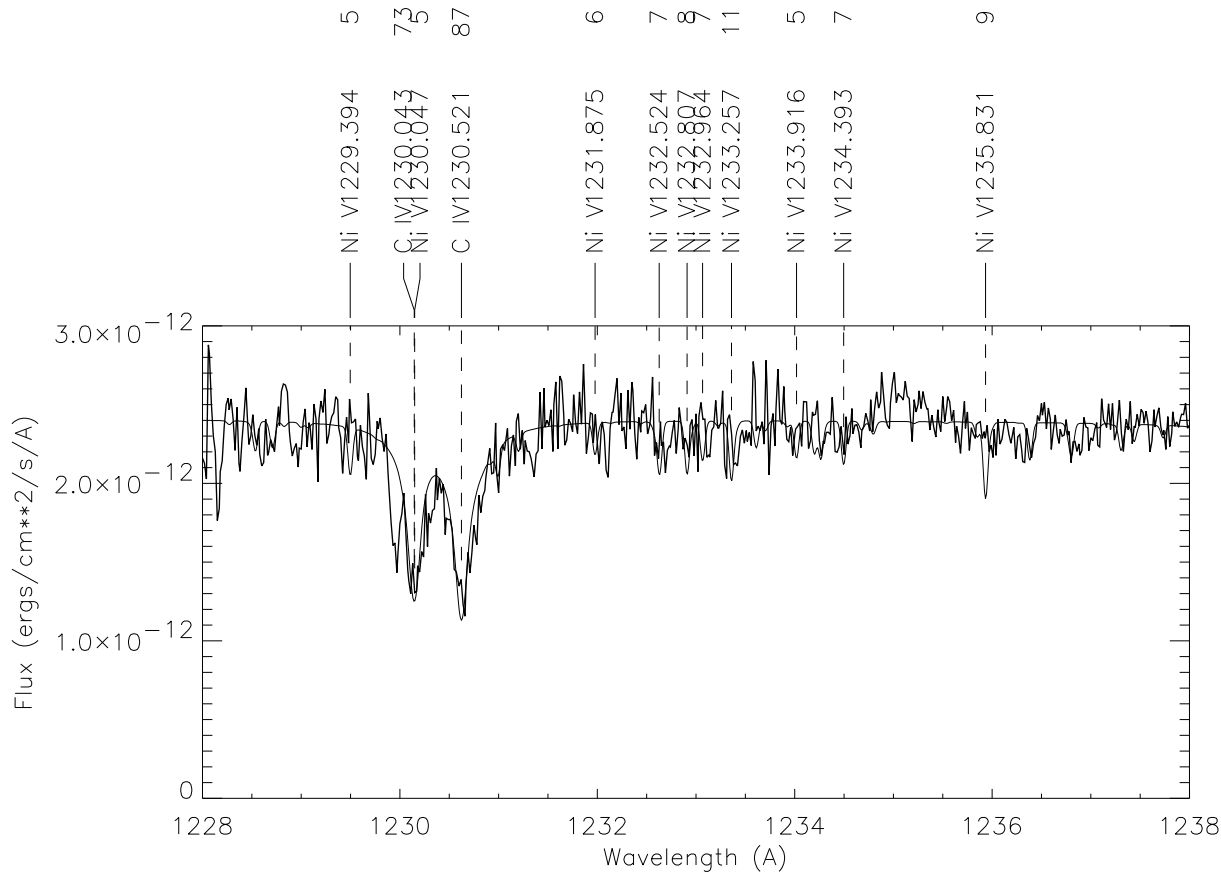}
\caption{a) GHRS spectra of REJ0503$-$289 after coaddition and merging of the 
individual exposures compared with a theoretical model. 
All photospheric spectral lines with predicted 
equivalent width above 5m\AA \ are marked, with their identification,
laboratory wavelength and predicted equivalent width (m\AA). Note
that the annotated equivalent widths are only approximate automated
determinations, whereas all values reported elsewhere (in the text
and tables) have been accurately measured from the data and synthetic
spectra.
The C, N, O and Si
abundances incorporated in the model were as listed in Table~\ref{mods}.
The Fe and Ni abundances chosen from the model grid were both $10^{-5}$.
The synthetic spectrum has been convolved with a 
0.042\AA \ (fwhm) gaussian function to represent the instrumental response.
}
\label{ghrs2}
\end{figure*}

\begin{figure*}
\leavevmode\epsffile{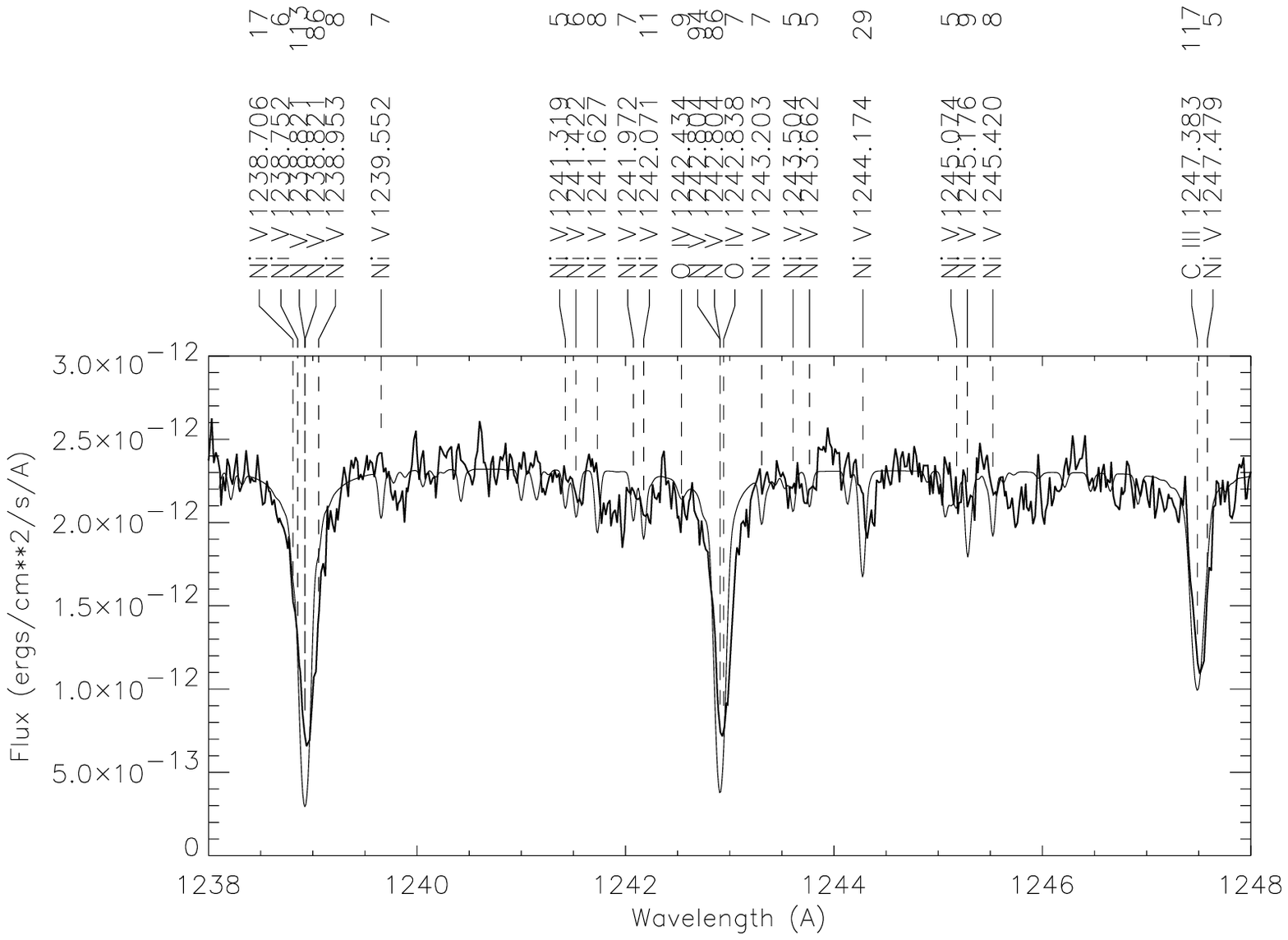}
\contcaption{b).}
\end{figure*}

\begin{figure*}
\leavevmode\epsffile{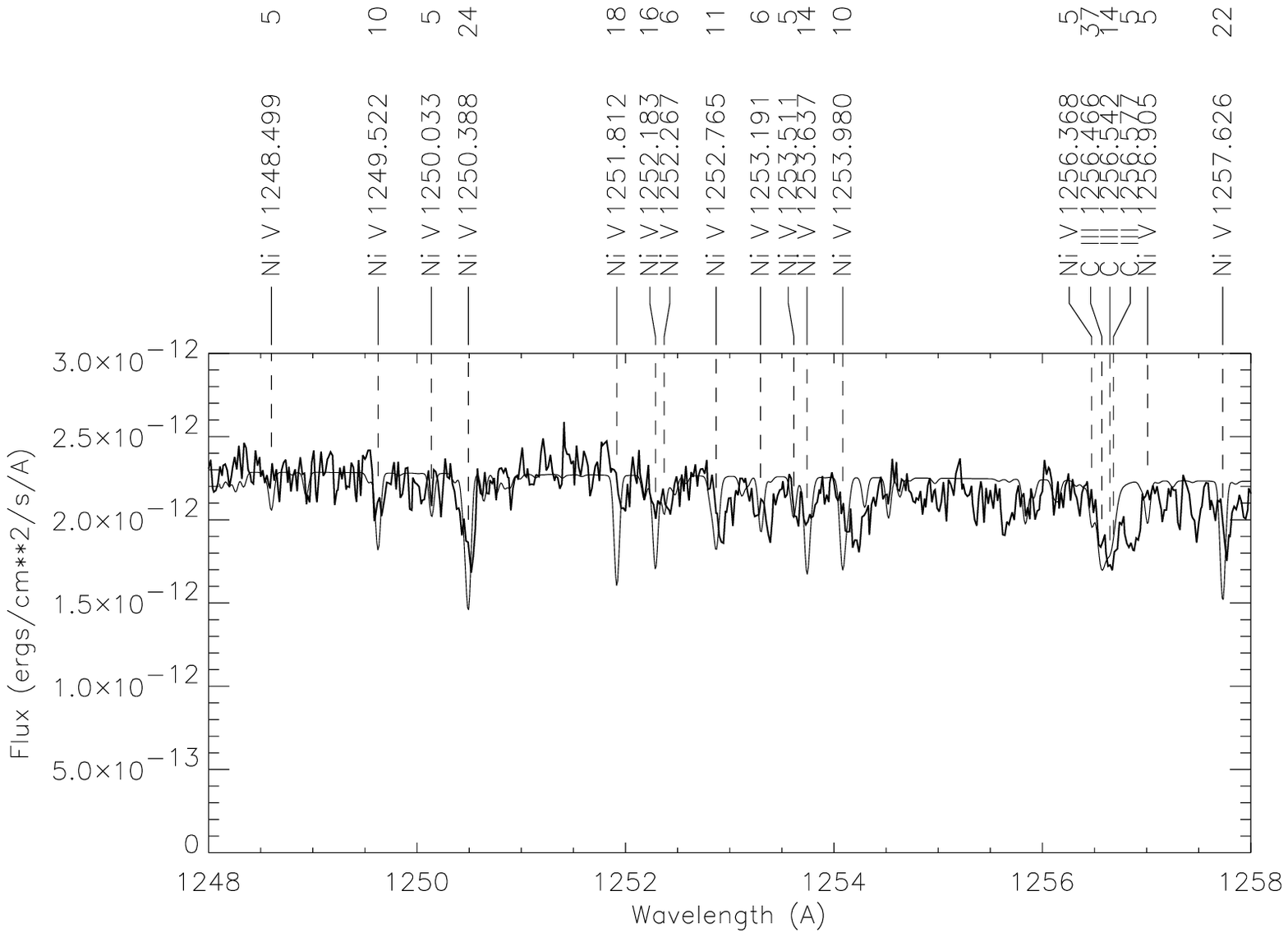}
\contcaption{c).}
\end{figure*}

\begin{figure*}
\leavevmode\epsffile{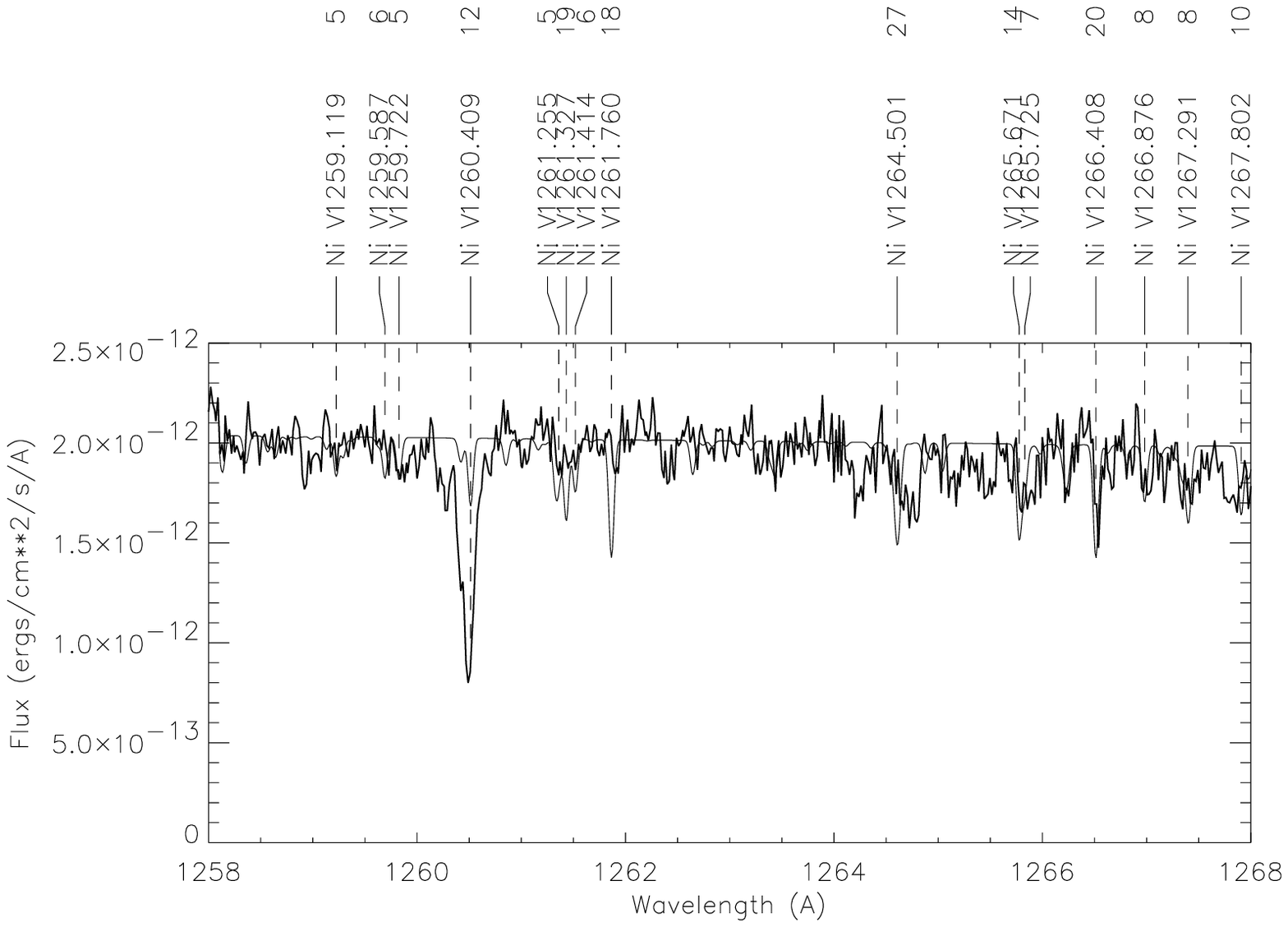}
\contcaption{d).}
\end{figure*}

\begin{figure*}
\leavevmode\epsffile{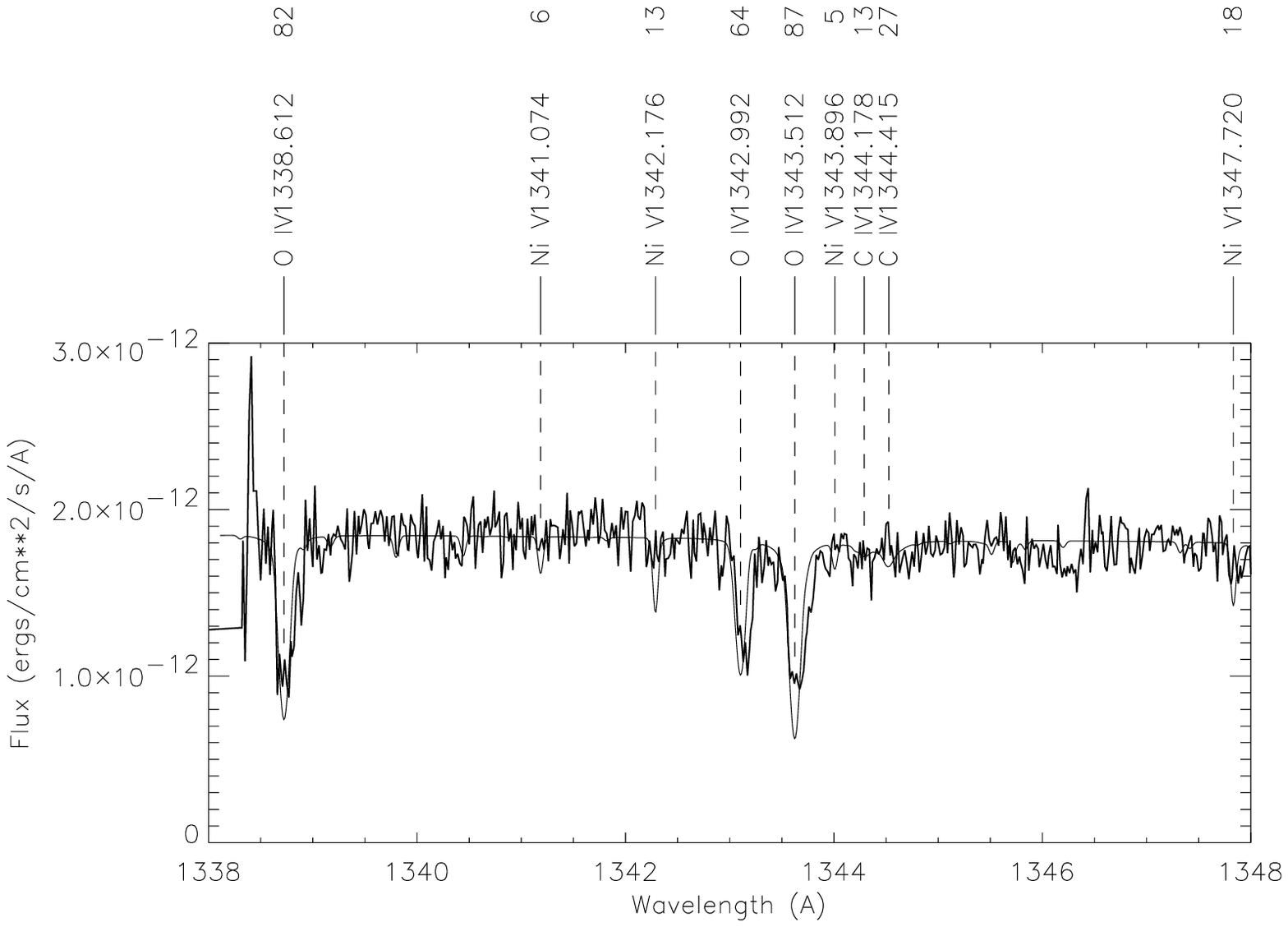}
\contcaption{e).}
\end{figure*}

\begin{figure*}
\leavevmode\epsffile{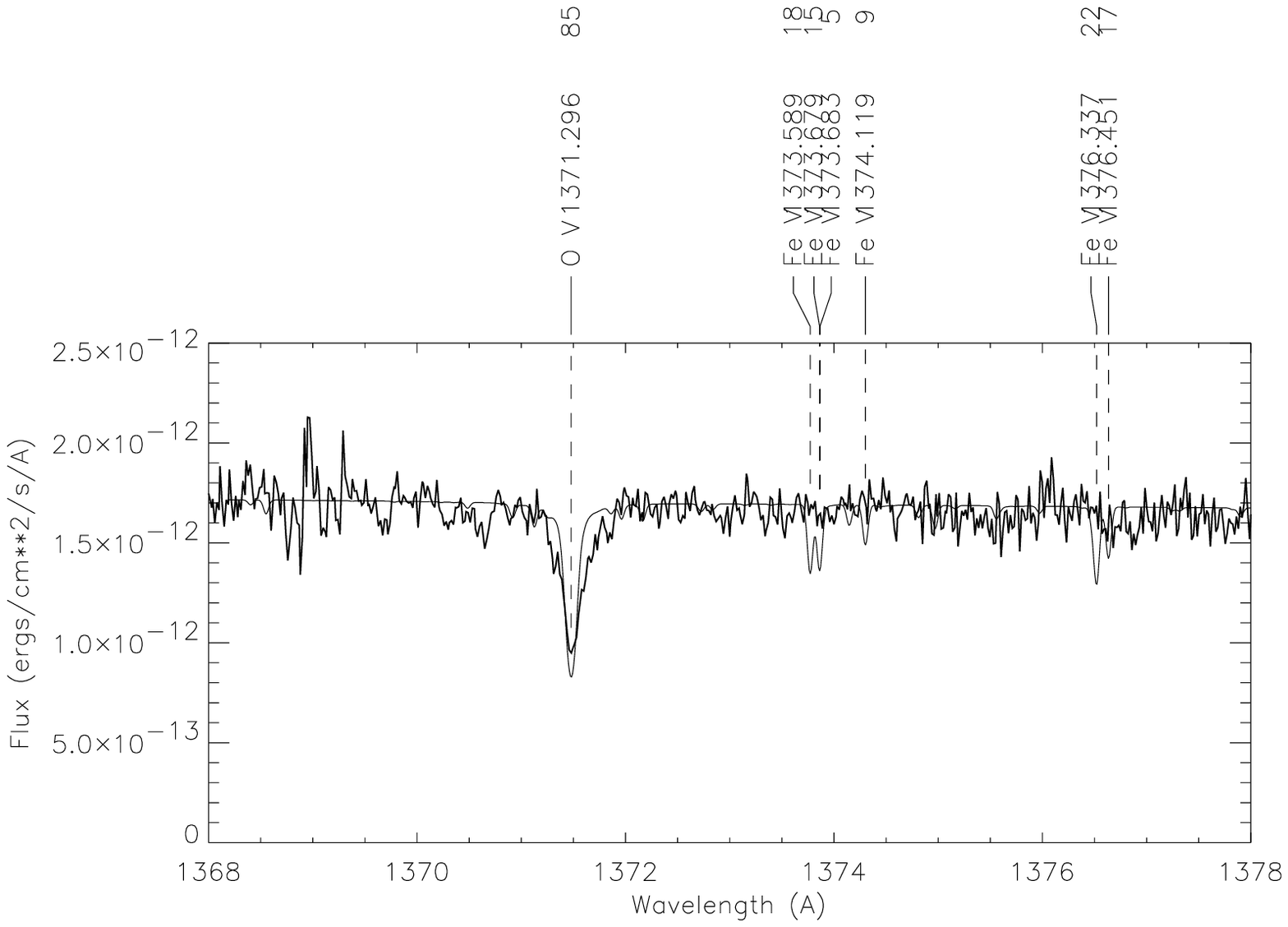}
\contcaption{f).}
\end{figure*}

\begin{figure*}
\leavevmode\epsffile{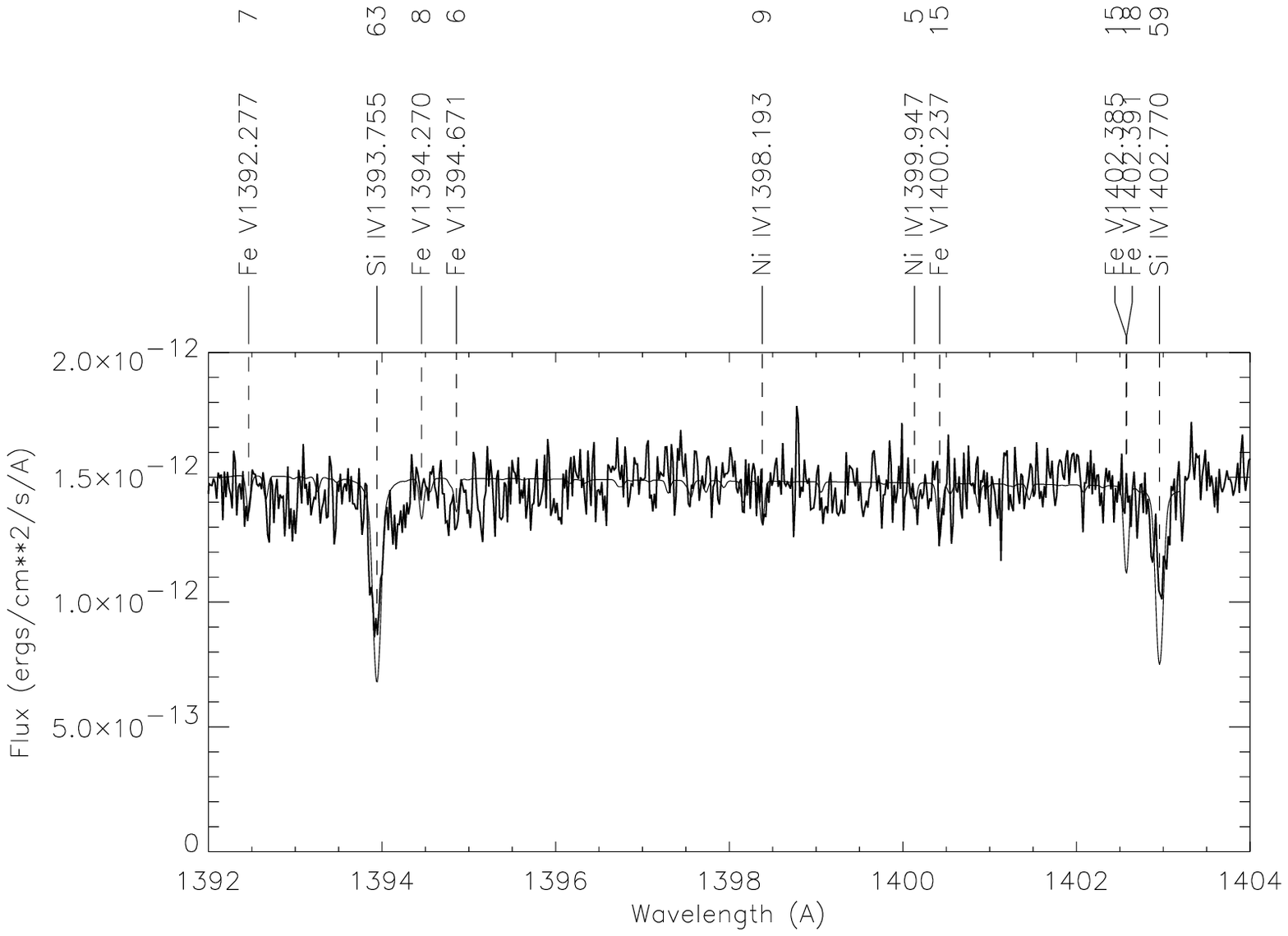}
\contcaption{g).}
\end{figure*}

\begin{figure*}
\leavevmode\epsffile{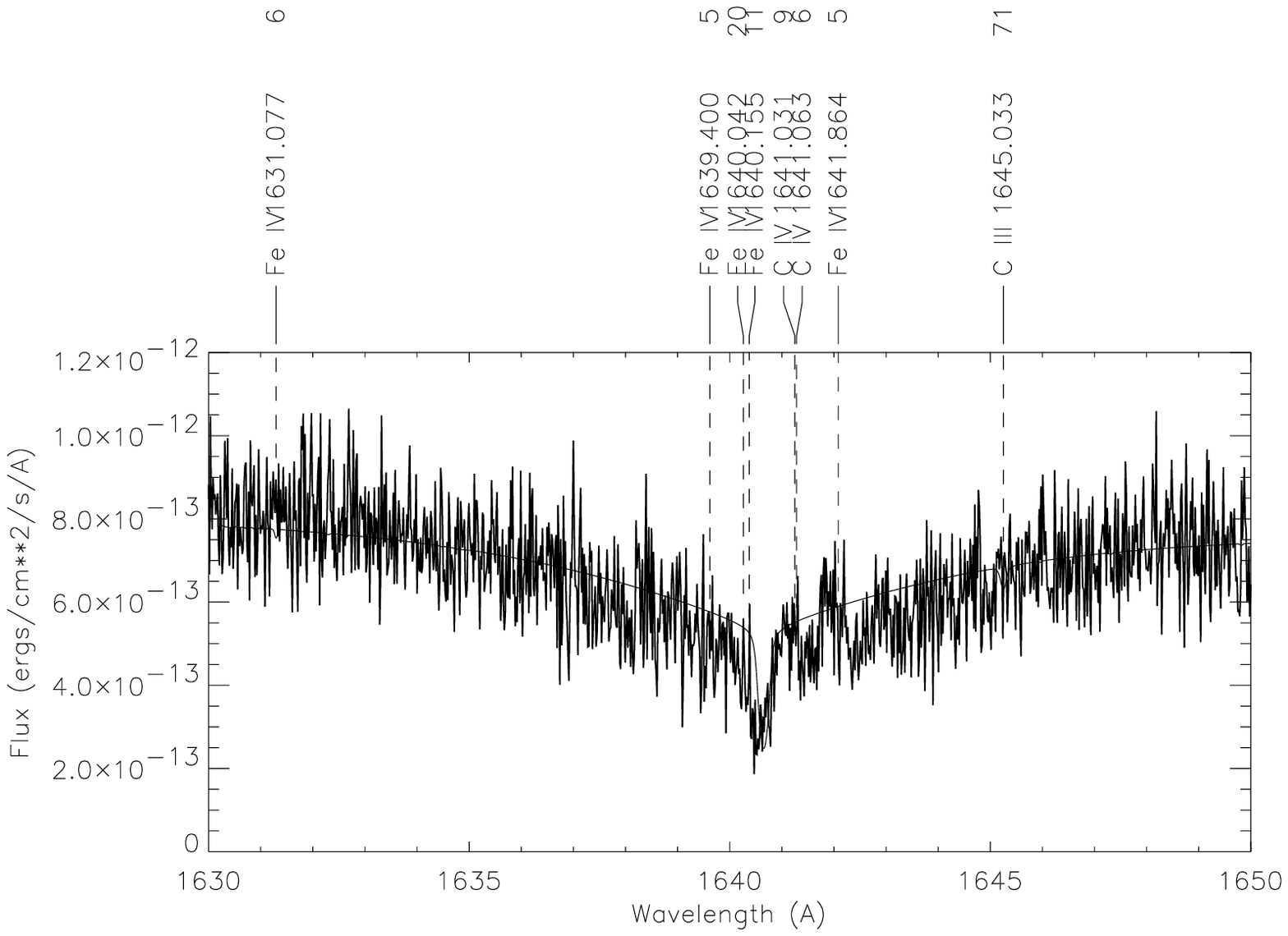}
\contcaption{h).}
\end{figure*}

\section{Observations}

\subsection{Optical spectrum}

The optical spectrum of REJ0503$-$289 was obtained by one of us (D.F.)
during an observational campaign aimed at the determination of the
temperature scale of DA white dwarfs (Finley, Koester \& \  Basri 1997). 
The spectrum of REJ0503$-$289 was obtained on
1992 September $22^{\rm nd}$ at the 3\,m Shane telescope of the
Lick Observatory,
using
the Kast double spectrograph. The spectrograph was
configured to cover the optical
wavelength range from 3300\AA \ up to 7500\AA \ in one exposure. A dichroic
mirror divides the beam at about 5500\AA. The blue spectrum was
recorded
with a 1200 x 400 Reticon CCD providing a resolution varying
from a little over 4\AA \ FWHM at 4000\AA \ to about 6\AA \ FWHM
at 5000\AA , with an
entrance slit width of $2"$. The red side is recorded separately on a
second 1200 x 400 Reticon CCD. Exposure times were set to obtain a
peak S/N of $\approx 100$ in the blue spectral region, with the actual
S/N ranging from 50 to 110. More details of the instrument set up as 
well as the data
reduction are
described by (Finley, Koester \& \ Basri 1997).

\subsection{GHRS on the Hubble Space Telescope}

The \hs \ far UV spectra used in this work were obtained from two
separate observing programmes carried out during cycle\,6 in 1996 and 1997 
by two of
us (Barstow and Werner), using the Goddard High Resolution Spectrograph
(GHRS) before its replacement during the second servicing mission.  All
spectra utilised the G160M grating, yielding a spectral resolution
$\approx 0.018$\AA \ rms, the grating angle adjusted to sample different
wavelength ranges.  The Barstow programme obtained 6 separate exposures
covering wavelengths 1233-1271\AA \ (2 spectra), 1369-1406\AA \ (3 spectra)
and 1619-1655\AA \ (1 spectrum). The main purpose of the multiple
exposures was to monitor any possible variation in absorption line
strengths that might be associated with a reported episodic wind (Barstow
\& \ Sion 1994). In contrast, the observations of Werner comprised just
two single spectra spanning the ranges 1225-1265\AA \ and 1335-1375\AA .
Table~\ref{obs} summarises all the GHRS observations.
Within
the observational errors, there is no evidence for any changes in the
profile of any of the \nv \  (1238.8/1242.8\AA ), \ov \  (1371.3\AA ) and \siiv \ 
(1393.8/1402.8) resonance lines. Measurement of the individual equivalent
widths (Table~\ref{linesw}) also fails to show any of the variability that
might be associated with the episodic wind reported by Barstow \& \ Sion
(1994).

\begin{table*}
\caption{Summary of GHRS observations of REJ0503$-$289}
\begin{tabular}{lcccl}
Date & Start time & Wavelength range & Exposure time & Observer \\
                & & (\AA ) & (s)\\
1996 November 23 & 07:34:34 & 1369-1406 & 653 & Barstow \\
                                 & 09:00:58 & 1233-1271 & 435 & \\
                                 & 09:10:59 & 1369-1406 & 544 & \\
                                 & 09:22:42 & 1619-1655 & 979 & \\
                                 & 10:45:44 & 1369-1406 & 544 & \\
                                 & 10:57:27 & 1233-1271 & 435 & \\
1997 January 03      & 09:22:30  & 1225-1265 & 761 & Werner \\
                     & 09:39:22  & 1335-1375 & 761 & \\
\end{tabular}
\label{obs}
\end{table*}

\begin{table*}
\caption{Equivalent widths (m\AA) of photospheric lines from
the multiple exposures. The measurement uncertainties are
in brackets.}
\begin{tabular}{lclllll}
Date & Start time & \nv \ 1238.3\AA & \nv \ 1242.2\AA & \ov \ 1371.3\AA & \siiv \ 
1393.6\AA & \siiv \ 1402.8\AA \\
1996 November 23 & 07:34:34 & & & 118.6(11.9) & 70.9(16.8) & 64.8(13.0)  \\
                                 & 09:00:58 & 219.7(17.3) & 164.8(15.3) \\
                                 & 09:10:59 & & & 109(13.3) & 78.5(14.1) & 61.1(15.2) \\
                                 & 10:45:44 && & 128.0(13.3) & 78.2(16.0) & 58.6(14.3) \\
                                 & 10:57:27 & 205.6(16.5) & 153.1(13.6)\\
1997 January 03  & 09:22:30      & 220.5(24.1) & 175.5(29.1)\\
                 & 09:39:22      &  && 124.6(21.4) \\
\end{tabular}
\label{linesw}
\end{table*}

With no apparent variability and repeated exposure of particular spectral
ranges, it is possible to generate a final data set with an overall
improvement in signal-to-noise, by merging and coadding the data in an
appropriate way. The procedure followed here was to first coadd the
results of the initial observations, where the spectral ranges are
identical for each exposure, weighting them according to their exposure
time. Secondly, the results of this exercise were merged with the later
observations, also weighted for exposure time, but only in the regions
where the data overlap. The resulting spectra are shown in
Fig.~\ref{ghrs2}.

\subsection{Temporal variability}

\begin{figure*}
\leavevmode\epsffile{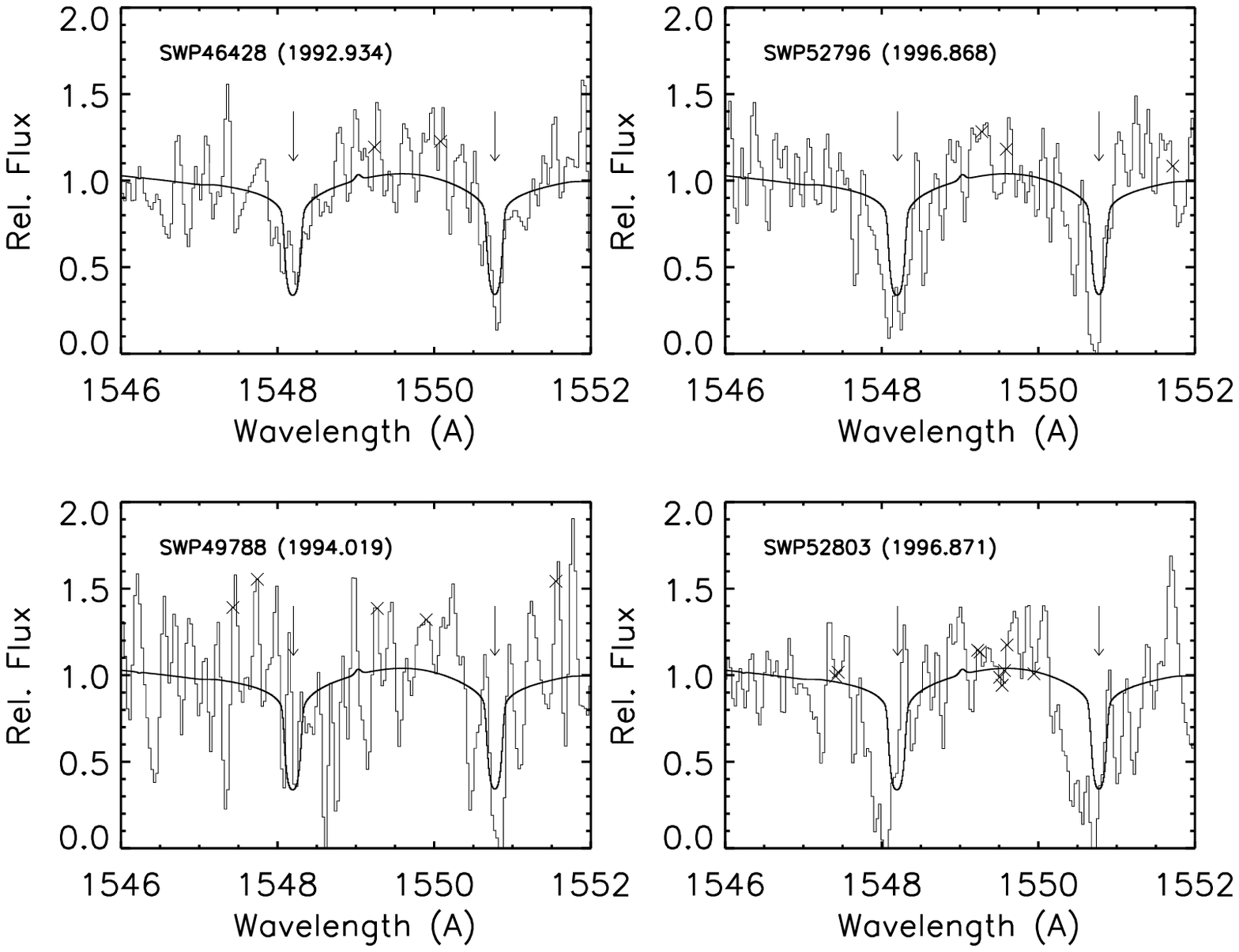}
\caption{A comparison of the region of the \civ \  resonance lines in the four
existing IUE echelle spectra of REJ0503$-$289
(histogram), illustrating temporal
changes in the line profiles. The location of the
resonance lines in the photospheric frame are indicated by arrows
and synthetic spectrum (smooth curve), calculated using the \tlus \ code is
shown for comparison.}
\label{var}
\end{figure*}

Although our GHRS spectra of REJ0503$-$289 show little evidence of
temporal variability such variations have been reported from \iu \  data.   As
previously mentioned, Barstow \& \ Sion (1994) reported evidence of
variations in the \civ , \ov \  and \heii \  features in two SWP 
echelle spectra
of REJ0503$-$289 obtained 13 months apart.  There exist two additional
spectra of this star, obtained subsequent to the Barstow \& \  Sion work, which
show further evidence of significant variations in the \civ \  resonance lines
(see Holberg, Barstow \& \ Sion 1998).  In Fig.~\ref{var} we show a comparison of
the region of the \civ resonance lines in all four spectra of REJ0503$-$289,
together with the predicted synthetic spectrum computed from a \tlus \ model. 
As is evident, the two spectra on the right, obtained in Nov. 1994, show
much more prominent \civ \ lines compared with those at the left, obtained
in Dec. 1992 and Jan. 1994.  In each spectrum a vertical line marks the
photospheric velocity of the star.  It appears that the most pronounced
change is associated with a strengthening of the blue wings of the C IV
lines and a possible development of a blue-shifted component
approximately 10 months later in 1994.  Similar blue-shifted components
are to be found in the majority of the hot He-rich degenerates observed
with \iu \  (Holberg, Barstow \& \ Sion 1998).  A discussion of the nature of
these blue-shifted features in DO stars is presented in Holberg, Barstow \& \ 
Sion (1999).

There is additional evidence of temporal variability in the comparison
between the equivalent widths as measured in the GHRS data and in the
IUE spectra.  Holberg, Barstow \& \ Sion (1998) presented results from a
coadded version of all four SWP spectra. These authors report equivalent
widths for the \nv \  resonance lines which are 30\% \ less than those in
Table~\ref{lines}. Such a change in equivalent width is significantly 
outside the
range of mutual uncertainties of the two data sets. The \siiv \  resonance
lines in both data sets, however,  remain consistent within mutual errors.

\section{Data analysis}

\subsection{Non-LTE model atmospheres}

We have used theoretical model atmospheres calculated using two
independent computer programmes in this study: the \tlus \ code developed
by Hubeny (Hubeny 1988; Hubeny \& \ Lanz 1992, 1995) and Werner's \pro \ 
suite (Werner 1986, Werner \& Dreizler 1999, Dreizler \& Werner 1993).  Both take account of non-LTE effects in the
calculations and include extensive line blanketing.

The \tlus \ models are an extension of work carried out on the atmospheres
of hot heavy element-rich DA white dwarfs by Lanz \etal \ (1996) and
Barstow \etal \ (1998) and have been described extensively in those
papers. Briefly, the models include a total of 26 ions of H, He, C, N, O,
Si, Fe and Ni. Radiative data for the light elements have been extracted
from TOPBASE, the database for the opacity project (Cunto \etal \ 1993),
except for extended models of carbon atoms. 
For iron and nickel, all the levels predicted by Kurucz
(1988) are included, taking into account the effect of over 9.4 million
lines.

To facilitate analysis of both the optical and far UV data sets, an
initial grid of models was calculated for determination of effective
temperature and surface gravity, spanning a range of $T_{\rm eff}$ from
65000K to 80000K (5000K steps) and for log g from 7.0 to 8.0 (0.5 dex
steps). For computation time reasons, these models only treated the
elements H, He and C. It is essential to deal explicitly
with carbon because it is
the only element, apart from He, visible in the optical spectrum. In
addition, the 1548/1550\AA \ resonance lines have a strong influence on
the temperature structure of the photospheric models and, as a result,
influence the \heii \  line strengths. The grid was extended to include the
heavier elements N, O, Si, Fe and Ni for the single value of log g=7.5,
but all values of $T_{\rm eff}$. Table~\ref{mods} lists the element
adundances included in the model grid.

\begin{table}
\caption{Element abundances included in \tlus \ model calculations, 
by number fraction with respect to helium. The Fe and Ni abundance which
are in best agreement with the observational data are in bold type.
For comparison, the final element abundances as derived with \pro \ models
are listed in the third column.
Where available the predicted abundances for $T_{\rm eff}=70000$K and
log g$=7.5$ are listed, from the radiative levitation calculations of
Chayer \etal \ (1995).}
\begin{tabular}{llll}
Element & Abundance & Abundance & Abundance\\
        & \tlus     & \pro      & rad. lev.\\
H       & $1\times 10^{-5}$ \\
He      & 1.0       & 1.0   \\
C       & 0.005     & 0.005     & 0.0002 \\
N       & $1.5\times 10^{-5}$ &$1.6\times 10^{-5}$ &$5\times 10^{-4}$ \\
O       & $3\times10^{-4}$ &$5\times10^{-4}$& $2\times 10^{-4}$\\
Si      & $10^{-5}$ &$10^{-5}$ & $<10^{-9}$\\
Fe      & ${\bf 1\times 10^{-6}}-3\times 10^{-5}$ &$< 5\times 10^{-6}$& $2\times 10^{-4}$\\
Ni      & ${\bf 1\times10^{-5}}-3\times 10^{-5}$&$1\times 10^{-5}$\\
\end{tabular}
\label{mods}
\end{table}

\pro \ is a code for calculating NLTE model atmospheres in radiative and
hydrostatic equilibrium using plane-parallel geometry. The \pro \ models
are an extension of the work of Dreizler \& \ Heber (1998). In addition to 
H, He, C, N, and O we included Fe and Ni in the same way as treated  by 
Werner, Dreizler \& \ Wolff (1995). In this set of model atmospheres we
used the previously determined parameters for $T_{\rm eff}$ and log g (Dreizler
\& Werner 1996), which took into account the strong influence of H, He and
C for the structure of the atmosphere due to the same reasons discussed
above. 

\begin{figure*}
\leavevmode\epsffile{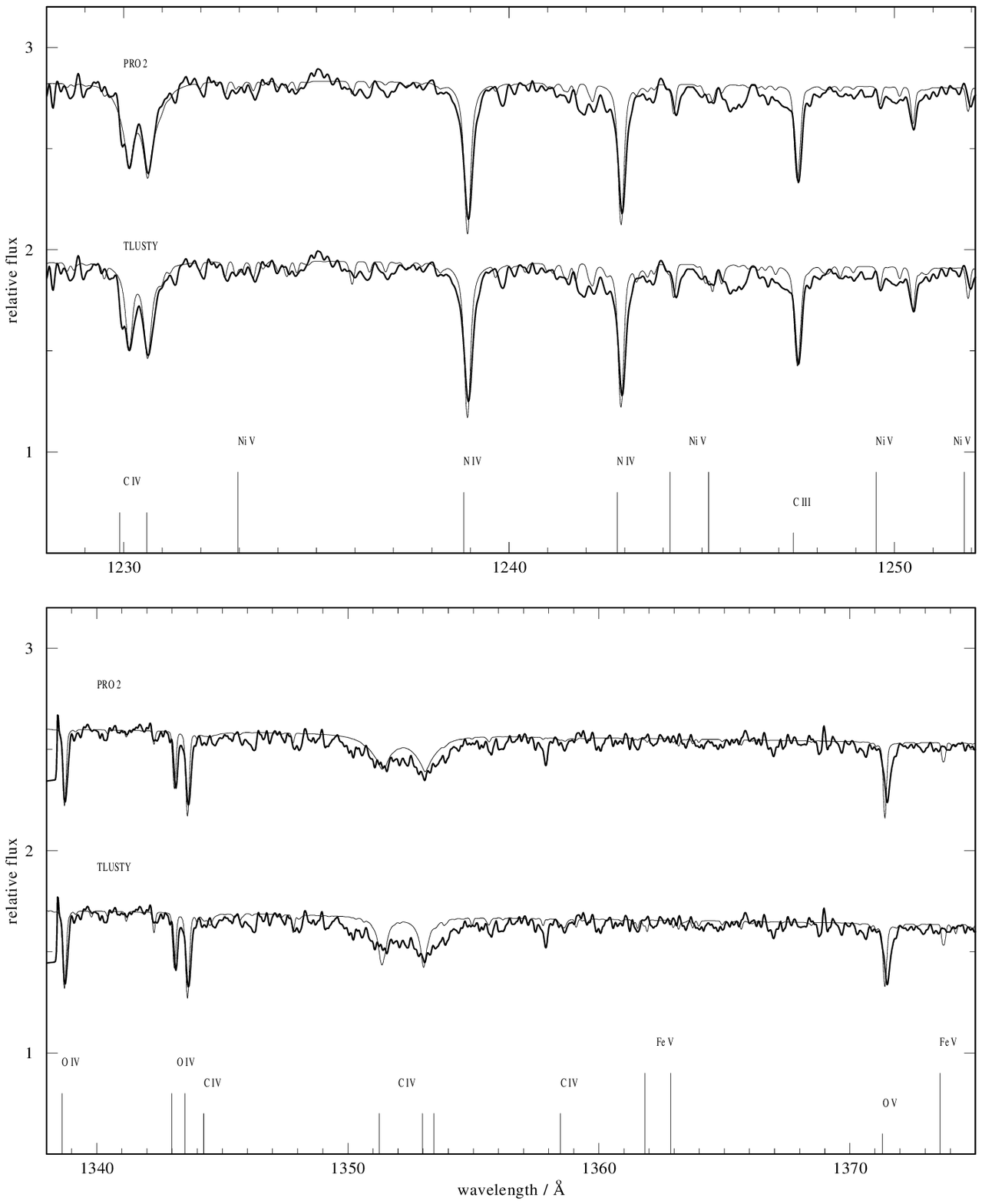}
\caption{Comparison of the \pro \ and \tlus \ model calculations (thin
curves).
1228\AA \ to 1252\AA \ (upper panel) and 1338\AA \ to 1375\AA (lower panel)
regions are shown and matched to the observational data (thick curves),
which have been smoothed using a 0.1\AA \ (fwhm) gaussian function to remove
some of the noise.
}
\label{pro2fit}
\end{figure*}

The physical
approximations as well as the atomic input data used in \tlus \ and
\pro \ are very similar but the code itself and the numerical techniques as 
well as the
treatment of the atomic data are completely independent. It is, therefore,
very interesting to compare the two sets of model atmospheres. Differences
allow a reliable estimation of the systematic errors. It is very satisfying
from the point of view of the modelers that these are below other
uncertainties like the flux calibration of the spectra. This is
demonstrated in Table~\ref{mods} and in Fig.~\ref{pro2fit}, where the
1228\AA --1252\AA \ and 1338\AA -- 1375\AA \ regions are compared with
synthetic spectra from \pro \ (top) and \tlus \ (bottom). The data
are smoothed by a 0.1\AA \ (fwhm) gaussian to reduce the noise in the observed
spectrum. 
It is interesting
to note, in Fig.~\ref{pro2fit}, that the line broadening approximation
adopted in \pro \ gives slightly better results than that used by
\tlus /\syn. 
Vennes \etal \  (1998) using
\iu \  and \orf \  spectra of REJ0503$-$289, together with LTE models,
determined abundances for C, N, and O which on average are over an
order of magnitude greater than those determined here. In their study,
only Si had a larger abundance, by a factor of 4, than our results.

\section{An optical determination of effective temperature and surface gravity}

The technique of using the H Balmer lines to estimate the effective
temperatures and surface gravities of the H-rich DA white dwarfs is
well-established (e.g. Kidder 1991; Bergeron Saffer
\& \ Liebert 1992). These measurements are made by comparing the observed
line profiles with the predictions of synthetic stellar spectra, computed
from theoretical model atmospheres, searching and interpolating a model
grid to find the best match.  An objective test, such as a $\chi ^2$
analysis, is used to determine the best-fit solution which also allows
formal determination of the measurement uncertainties. While some
questions have been raised about the limitations of the technique for the
highest temperature DA white dwarfs, with heavy element contaminated
atmospheres and weak Balmer lines 
(Napiwotzki 1992, Napiwotzki \& \ Rauch 1994,
Barstow Hubeny \& \ Holberg 1998), it
remains the most important and widely used means of determining the 
DA temperature scale.

In contrast, it has been more difficult to establish a similar standard
technique for the DO white dwarfs. This has partly arisen from problems in
obtaining good agreement between the models and the data. In particular,
it has been difficult to find models that can match all observed \heii \ 
profiles simultaneously (e.g. see Dreizler \& \ Werner 1996).  A
particular problem has usually been obtaining a good fit to the $\lambda
4686$ line. Consequently, in determining $T_{\rm eff}$ and log g for their
sample of DO white dwarfs (the first comprehensive survey), Dreizler \& \
Werner (1996) were forced to rely on a simple visual comparison to select
the best-fit model. There is no reason to suppose that the parameters
determined for REJ0503$-$289 and the other DOs studied are seriously in
error. However, it is certainly difficult to determine the possible
uncertainties in the measurements. Here we develop a more objective
approach to obtaining $T_{\rm eff}$ and log g by fitting the He lines present
in the optical spectrum of REJ0503$-$289 in a manner similar to the
technique applied to the DA white dwarf Balmer lines.

\begin{figure*}
\leavevmode\epsffile{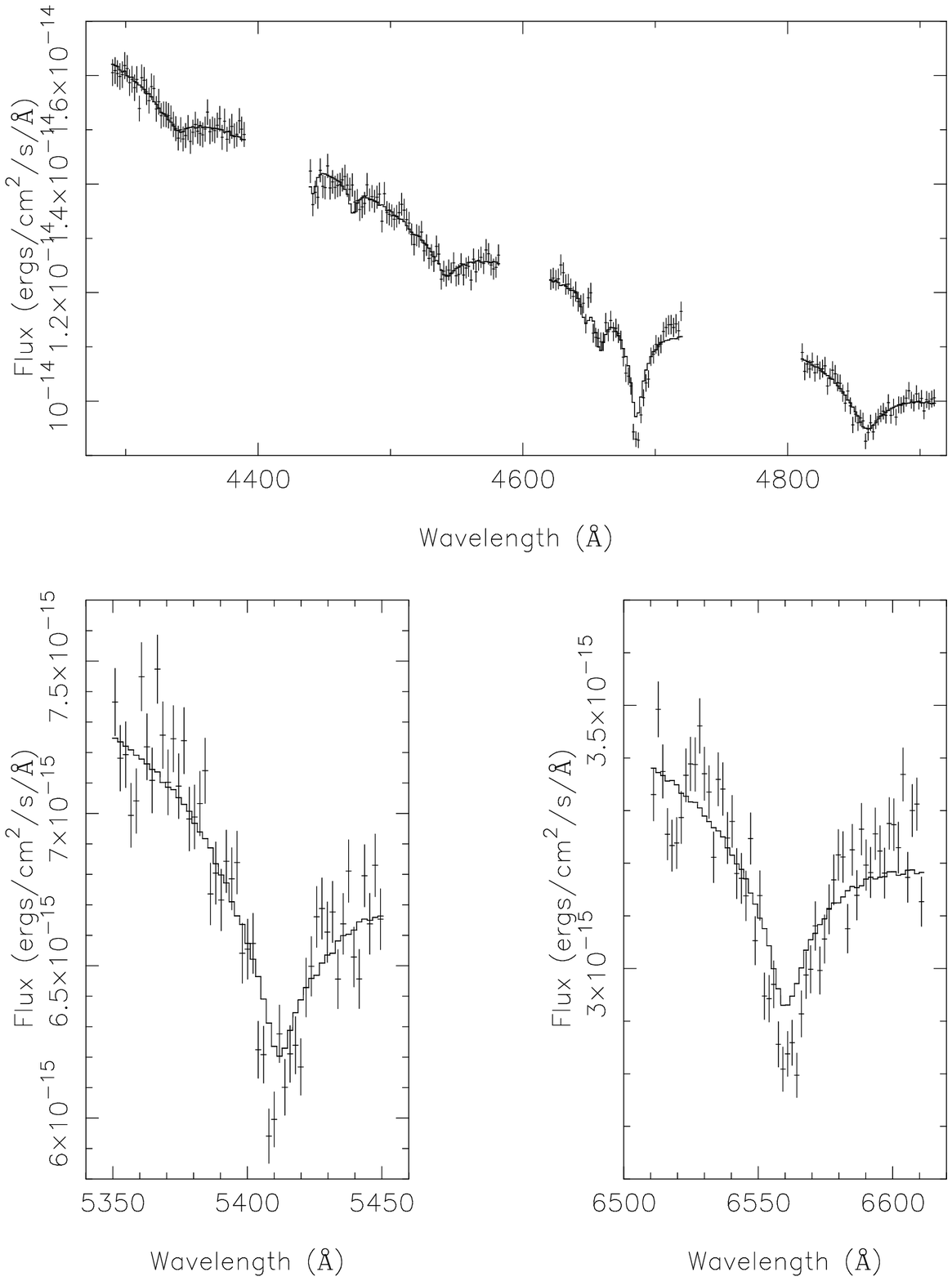}
\caption{Comparison of the observed helium lines (error bars) with the
predictions of a synthetic spectrum computed from a model atmosphere
including He, H and C. The lines selected for this analysis are (from top
to bottom of the figure) listed in Table~\ref{lfit}.
Top panel -- \heii \ 4339, \hei/\heii \  4471/4542,
\civ /\heii \ 4659+4647/4686, \heii \  4860; bottom left panel --
\heii \ 5400; bottom right panel -- \heii \ 6560.
}
\label{optfit}
\end{figure*}

Several \heii \  lines are visible in the optical spectrum of REJ0503$-$289
(Fig.~\ref{optfit}), from 4339\AA \ to 6560\AA . In addition, there are
important \hei \  lines at 4471\AA \ and
5876\AA , not detected in the spectrum, which are
very sensitive to effective temperature and, therefore, should be
included in any analysis. However, for this spectrum, the
5876\AA \ line falls in a region of comparatively noisy data, providing
a weaker constraint than the 4471\AA \ feature. Hence, we only consider the
latter line in this analysis.  
For this paper, we have adapted the technique
used in our previous work (e.g. Barstow \etal \ 1998), splitting the
spectrum into discrete regions spanned by the absorption lines for
comparison with the synthetic spectra. The features included and the
wavelength ranges needed to capture the complete lines are listed in
Table~\ref{helin}. The complex \civ /\heii \  blend is covered in a single
section of data and the \hei \  4471\AA \ feature is incorporated with the
\heii \  4542\AA \ line.

\begin{table}
\caption{List of He lines used for determination of $T_{\rm eff}$ and log g, 
indicating the central wavelength ($\lambda_c$), wavelength range 
($\Delta \lambda$), identified lines and their rest 
wavelengths ($\lambda_r$) considered in the analysis.}
\begin{tabular}{ccll}
$\lambda_c$ (\AA) & $\Delta \lambda$ (\AA) & Species &$\lambda_r$
(\AA) \\
4339 & 100 & \heii \  & 4339 \\
4542 & 140 & \hei/\heii \  & 4471/4542 \\
4670 & 100 & \civ /\heii \  &  4659+4647/4686 \\
4860 & 100 & \heii \  & 4860\\
5400 & 100 & \heii \  & 5411\\
6560 & 100 & \heii \  & 6560\\
\end{tabular}
\label{helin}
\end{table}

We used the programme \xsp \ (Shafer \etal \ 1991) to compare the
spectral models with the observational data. \xsp \ utilises a robust
$\chi^2$ minimisation routine to find the best match to the data.  All
the lines included were fit simultaneously and an independent
normalisation constant was applied to each, reducing the effect
of any wavelength dependent systematic errors in the flux calibration 
of the spectrum. \xsp \
interpolates the synthetic spectra linearly between points in the model
grid. Any wavelength or velocity
shifts were accounted for by allowing the radial velocity of the lines
to vary (taking identical values for each line) during the fit. Once, the
best match to the velocity had been obtained, this parameter was fixed, being
of no physical interest in this work. The carbon abundance was fixed at
a single value of C/He$=0.005$ throughout the analysis.
Provided the model corresponding to minimum $\chi^2$ can be
considered to be a `good' fit ($\chi^2_{red}<2$: $\chi^2_{red}=\chi^2/\nu$,
where $\nu $ is the number of degrees of freedom), the uncertainties in
$T_{\rm eff}$ and log g can be determined by considering the departures in
$\chi^2$ ($\delta \chi^2$) from this minimum. 
For the two parameters of
interest in this analysis (the variables $T_{\rm eff}$ and log g), a $1\sigma
$ uncertainty corresponds to $\delta \chi^2=2.3$ (see Press \etal \ 1992).
It should be noted that this only takes account of the statistical
uncertainties in the data and does not include any possible systematic
effects related to the model spectra or data reduction process.

Fig.~\ref{optfit} and Table~\ref{lfit} show the good agreement achieved
between the best fit model and the data. The value of $\chi^{2}_{red}$
(1.49) is clearly indicative of a good match between model and data.
However, inspection of the fit to the \heii \  4686\AA \ line shows that the
predicted line strength is weaker than observed, particularly in the line
core. If the 4686\AA \ line is ignored, the agreement between model and
data improves, lowering $\chi^2_{red}$ to 1.38 (see Table~\ref{lfit}).
There is also an increase in the estimated temperature, from 72660K to
74990K, and a slight lowering of the surface gravity (by 0.1 dex).

It has recently been reported by Barstow \etal \ (1998) that the presence
of significant quantities of heavy elements, in particuar Fe and Ni, in
the atmospheres of DA white dwarfs has a significant effect on the temperature
scale. For example, a lower value of the effective temperature is
measured from the Balmer line profiles when these elements are included
self consistently in the model calculations, in comparison with the
results of pure H or H+He models. In the analysis of the He-rich star
REJ0503$- $289, discussed here, we included a significant abundance of
carbon but no other heavy elements in the theoretical models. However, it
is known from earlier work (Barstow \& \ Sion 1994; Barstow \etal \ 1996)
that O, N and Si are definitely present, although at lower abundances
(with respect to He) than C, and there were also hints at the possible
presence of Fe and Ni.

To test the possible effect of these other heavy elements we extended the
spectral grid, fixing log g at a single value of 7.5 but covering the original
65000K to 80000K temperature range. The nominal heavy element abundances
incorporated are listed in Table~\ref{mods}. To assess the possible impact
of treating all the heavy elements on the determination of effective
temperature, the He lines analysis was repeated using the full heavy
element models but with fixed values of the Fe/He and Ni/He abundances
($10^{-5}$ in both cases). An $\approx 4$\% decrease in the
effective temperature is seen, but it should be noted that the
experimental error bars overlap considerably. 

\begin{table*}
\caption{Results of the optical spectral analysis carried out using 
the programme \xsp , listing the best-
fit temperature and  surface gravity together with the $1\sigma $ 
uncertainties.}
\begin{tabular}{llll}
Parameter & Fit to all lines & Fit excluding \heii \  4686\AA 
& Fit to all lines (heavy element models)\\
$\chi^2$ & 508.1 & 398.5 & 492.9\\
$\chi^2_{red}$ & 1.49 & 1.38  & 1.45 \\
$T_{\rm eff}$ & 72660 (66371--75613) & 74990 (73150--76630) & 70000
(69390--70650)\\
log g & 7.50 (7.35--7.63) & 7.41 (7.26--7.53) & 7.5 fixed\\
\end{tabular}
\label{lfit}
\end{table*}

\section{Determination of heavy element abundances from the GHRS data}

Since the effective temperature and surface gravity of REJ0503$-$289 are
close to 70000K and 7.5 respectively, one of the points of the model
atmosphere grid, we adopt these values when applying model calculations
to this particular analysis. Fig.~\ref{ghrs2} shows the merged GHRS
spectrum, described in section 2.2, together with a synthetic spectrum
computed for the nominal C, N, O and Si abundances listed in
Table~\ref{mods} and with the Fe and Ni abundances fixed at $10^{-5}$.
The synthetic spectrum has been convolved with a 
0.042\AA \ (fwhm) gaussian function to represent the instrumental response.
The positions of all lines with a predicted equivalent width greater or
equal to 5m\AA \ are marked. Several strong lines from highly ionized
species of C, N, O and Si are clearly visible, which are most likely to
be of photospheric origin. For the most part these have already been
identified in the \iu \ echelle spectra of this star (Barstow \& \ Sion
1994; Holberg Barstow and Sion 1998) but we list them here, together with
their rest wavelengths, measured wavelengths and measured equivalent
widths (Table~\ref{lines}). However, the improved signal-to-noise of the
GHRS spectrum, compared to even the coadded \iu \ data allows the
detection of several new features, which are included in
Table~\ref{lines}. Another important factor may also be the absence of any
so-called reseau marks in the GHRS, an important feature of the \iu \
spectra arising from the spatial calibration. Also visible is a single
interstellar line of \siii \  at 1260.4221\AA .

\begin{figure*}
\leavevmode\epsffile{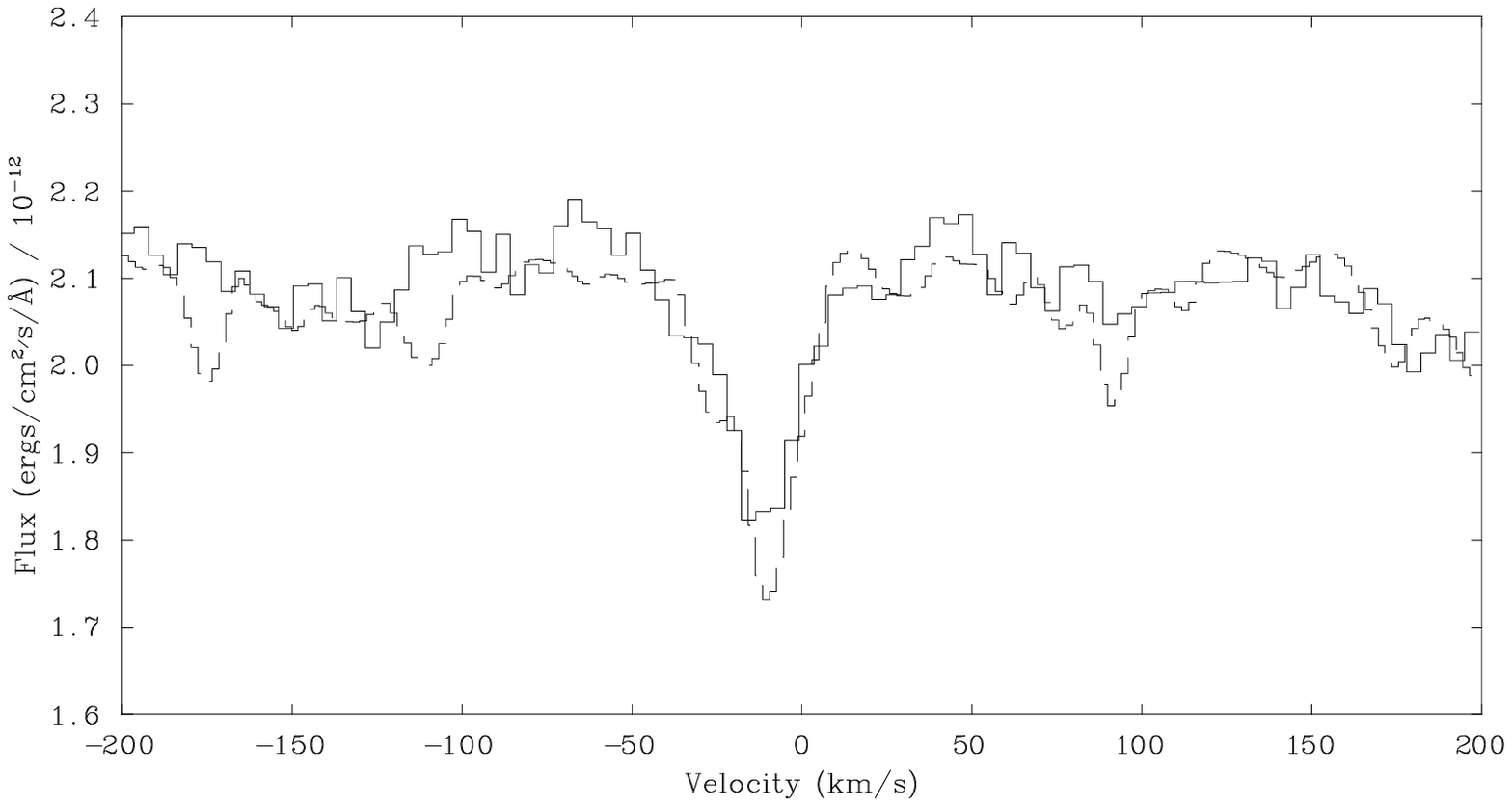}
\caption{The eight \niv \  lines predicted to be strongest in the GHRS spectrum 
coadded in velocity space (solid histogram). A synthetic spectrum, 
calculated from a model containing a Ni/He ratio of  $10^{-5}$ (dashed 
histogram) is shown for comparison. All other elemental abundance are as 
listed in Table~\ref{mods}.}
\label{nivel}
\end{figure*}

Apart from these very strongest lines, which are clearly detected, any
other possible absorption features are close to the limits of detection
imposed by the general signal-to-noise of the data. However, further
inspection of Fig.~\ref{ghrs2} shows a number of coincidences between
possible features and predicted \niv \  lines. Several of these observed
features, at 1250.4\AA , 1257.6\AA \ (Fig.~\ref{ghrs2}c)
and 1266.4\AA \ (Fig.~\ref{ghrs2}d), are almost strong
enough to constitute detections in their own right. On the basis of these
features alone, the evidence for the presence of Ni in the photosphere of
REJ0503$-$289, is not very strong. As a further test we coadded the eight
\niv \  lines predicted to be the strongest ($\lambda \lambda$1244.23, 1250.41, 
1252.80, 1253.25,
1254.09, 1257.66, 1264.62, 1266.40) in velocity space
(Fig.~\ref{nivel}). This technique has been applied to \iu \ data in
the past to detect elements such as Fe and Ni, which do not have any
particularly strong resonance transitions but large numbers of
comparatively weak features (see e.g. Holberg \etal \ 1994). The
procedure involves shifting the spectra (in this case the GHRS data) into
a velocity frame of reference centred on the wavelength of a particular
line and then summing and averaging the results for several lines. If
apparent weak features are just random fluctuations of the noise the
coaddition process will tend to eliminate them whereas, if the features
are real, summing the spectra will produce a more significant combined absorption
line. The typical experimental uncertainties on the original coadded GHRS
spectrum are $\approx 7-10$\% , whereas the scatter from data point to
data point on the velocity coadded spectrum is around 2\%.

\begin{figure*}
\leavevmode\epsffile{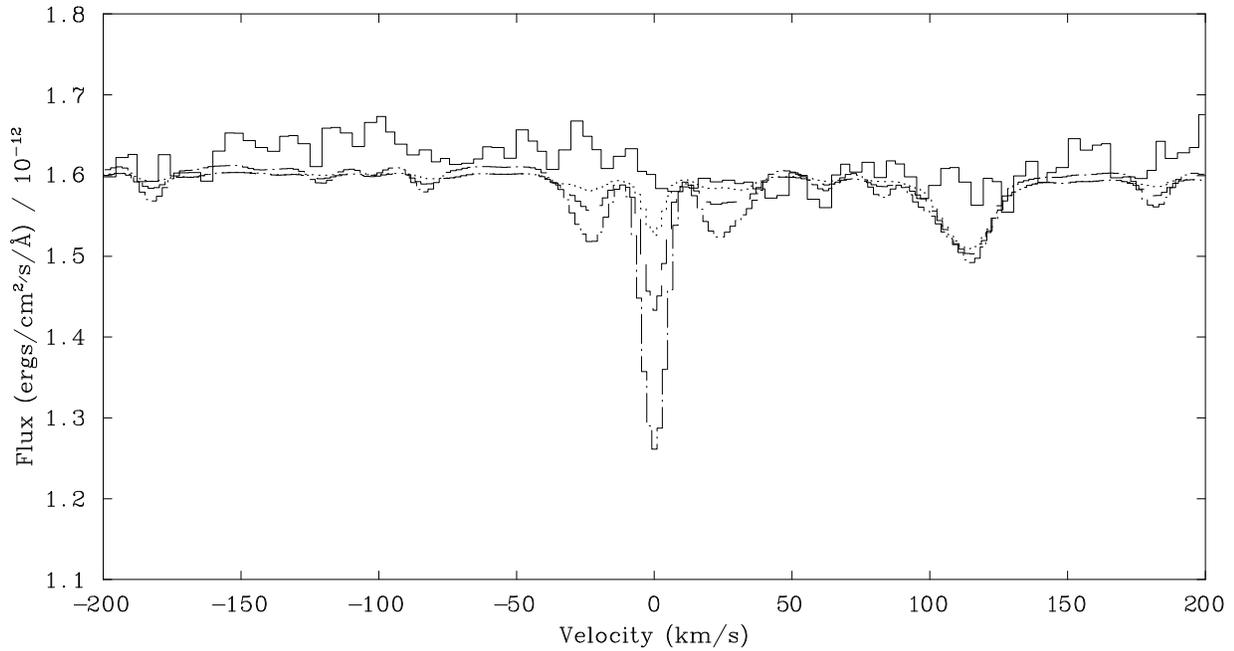}
\caption{The nine \fev \  lines predicted to be strongest in the GHRS spectrum 
coadded in velocity space (solid histogram).  Three synthetic spectra, 
calculated from models containing Fe/He ratios of $10^{-5}$, $3\times 
10^{-6}$ and $10^{-6}$ (dot-dashed, dashed and dotted histograms, in order of decreasing line 
strength) are  shown for comparison. All other elemental abundance are 
as listed in Table~\ref{mods}.}
\label{fevel}
\end{figure*}

Fig.~\ref{nivel} shows the results of the coaddition of the 8 Ni lines.
For comparison, the same procedure was carried out for the synthetic
spectrum. An absorption feature is clearly detected in both the data and
model, at approximately the same strength. This is clear evidence that Ni
is present in the photosphere of REJ0503$-$289 at an abundance of
$\approx 10^{-5}$ with respect to He, although the predicted, coadded line
strength is a little stronger than the observation. Interestingly, while
there are fewer Fe lines expected to be present in these spectral ranges
and their predicted equivalent widths are typically smaller than those of
the Ni lines, there are no similar coincidences where significant Fe
lines are expected. However, the constraints placed on the Fe abundance
by individual lines are not particularly restrictive, only implying an Fe
abundance below $\approx 10^{-5}$
(see Fig.~\ref{ghrs2}f). Again, coaddition of the nine
strongest predicted Fe lines ($\lambda $1361.826, 1373.589, 1373.679,
1376.337, 1376.451, 1378.561, 1387.095, 1387.937, 1402.237) provides a
more sensitive indication as to whether or not there is Fe present in the
atmosphere (Fig.~\ref{fevel}). However, Fe is not detected in this
case, as there is no sign of any absorption feature. Comparing the
coadded spectrum with models calculated for a range of Fe abundances from
$10^{-6}$ to $10^{-5}$ (Fig.~\ref{fevel}), gives an improved lower
limit to the Fe abundance $\approx 10^{-6}$.

\begin{table*}
\caption{Absorption lines detected in the GHRS spectrum of REJ0503$-$289.
The figures in brackets are the equivalent widths predicted by a spectral
model incorporating the nominal photospheric abundances (see
Table~\ref{mods}). No measurements are listed for the 
\civ \ lines at 1351.214\AA , 1351.286\AA \ and 1352.920\AA \ due to their
broad and shallow nature, making the core difficult to locate.
Identifications marked (?) are tentative.}
\begin{tabular}{llllllll}
Species & Lab. $\lambda $(\AA )& Obs.$\lambda $(\AA ) & Error &
V (km/s) & Error & EW (m\AA ) & Error \\
Interstellar\\
\ci \  (?)  & 1239.780 &   1239.822 &  0.056 &   10.21 & 13.53 &    22.9       & 10.65\\
\siii \     & 1260.422 &   1260.464 &  0.019 &    9.96 &  4.44 &   112.5       & 10.98\\
\\
Photospheric\\
\civ \      & 1230.043 &   1230.181 &  0.016 &   33.56 &  3.87 &    64.9 (72)  &  8.6\\
\civ \      & 1230.521 &   1230.644 &  0.019 &   30.07 &  4.57 &   118.4 (87)  & 11.2\\
\nv \       & 1238.821 &   1238.936 &  0.026 &   27.83 &  6.36 &   215.5 (199) & 15.2\\
\nv \       & 1242.804 &   1242.918 &  0.018 &   27.60 &  4.44 &   155.7 (200) & 11.9\\
\ciii \     & 1247.383 &   1247.509 &  0.013 &   30.37 &  3.06 &    79.0 (117) &  8.2\\
\niv \      & 1250.388 &   1250.479 &  0.019 &   21.98 &  4.65 &    29.1 (26)  &  7.8\\
\ciii \     & 1256.500 &   1256.627 &  0.027 &   30.38 &  6.40 &    31.6 (56)  &  9.0\\
\svi \  (?) & 1256.802 &   1256.868 &  0.031 &   15.81 &  7.35 &    15.9       &  7.5\\
\oiv \      & 1338.612 &   1338.764 &  0.006 &   33.96 &  1.43 &   100.7 (82)  &  6.3\\
\oiv \      & 1342.992 &   1343.148 &  0.010 &   34.77 &  2.18 &    58.6 (67)  &  6.9\\
\oiv \      & 1343.512 &   1343.658  & 0.009 &   32.58 &  2.05 &    90.9 (87)  & 7.2\\
\civ \      & 1351.214 \\
\civ \      & 1351.286 \\
\civ \      & 1352.920 \\
\ov \       & 1371.296 &   1371.497 &  0.020 &   43.95 &  4.38 &   114.6 (84)  & 11.4\\
\ciii \     & 1381.652 &   1381.851 &  0.041 &   43.07 &  8.85 &    53.1 (71)  & 12.5\\
\siiv \     & 1393.755 &   1393.932 &  0.018 &   38.10 &  3.76 &    50.7 (63)  &  8.5\\
\siiv \     & 1402.770 &   1402.964 &  0.014 &   41.35 &  2.88 &    46.2 (59)  &  7.5\\
Mean    &           &             &         &   33.54 &  6.09 \\
\end{tabular}
\label{lines}
\end{table*}

\section{Discussion}

The availability of GHRS spectra of REJ0503$-$289, coupled with a new
optical spectrum of the star has revealed important information regarding
the structure and evolution of this interesting DO white dwarf. We have
used the optical spectrum to determine the temperature and surface
gravity of the star which is broadly in agreement with earlier
determinations from the original ESO optical observation (Barstow \etal \
1994; Dreizler \& \ Werner 1996) and the \orf \ far-UV spectrum published
by Vennes
\etal \ (1998).

Perhaps the most important part of this particular optical analysis is the more
objective determination of the values of $T_{\rm eff}$ and log g, using a
spectral fitting technique, and their respective errors.  Whether or not
these results might be considered to be in agreement or disagreement with
the results of Vennes
\etal \ (1998) depends somewhat on how we choose to make the measurement, with
or without the \heii \  4686 line and using models with or without the
elements heavier than H, He or C. In fact such a comparison is probably
not particularly instructive as Vennes \etal \ used LTE models (compared
to our non-LTE calculations) including only H and He. What is important
is that we find that the inclusion of heavy elements in the models may
have an
influence on the outcome of the temperature determination, as it
does for the hot DA white dwarfs. Significant abundances of N, O, Si, Fe
and Ni (abundances estimated from the GHRS spectra), in addition to the H,
He and C treated in the simpler models, lower the value of $T_{\rm eff}$ by
approximately 2500K. Nevertheless, a detailed
study of any DO star probably needs to be entirely self-consistent,
combining temperature/gravity and abundance determinatons using spectra
from at least visible and UV wavelength ranges. We note that, in this
analysis, a higher Fe/He abundance than observed was included in the
models used. Therefore, the observed $T_{\rm eff}$ change should only
be regarded as an upper limit and it must be remembered that systematic
errors may be of similar magnitude.

Detections of nitrogen, oxygen and silicon in the far UV spectra have
already been reported by other authors (e.g. Barstow \etal \ 1996; Barstow
\& \ Sion 1994; Dreizler \& \ Werner 1996), while carbon is clearly seen
in both UV and visible bands. However, while the \iu \ data may have
hinted at the presence of Fe and/or Ni (see Barstow \etal \ 1996), we are
able to demonstrate that the star really does contain significant
quantites of nickel for the first time, using the GHRS data. This is
revealed initially in marginal detections of the strongest individual Ni
lines but clearly confirmed when the eight strongest Ni lines are coadded
in velocity space. This is the first detection of Ni in a non-DA white
dwarf.

Nickel has also been observed in a number of very hot H-rich DA white
dwarfs (e.g. Holberg \etal \ 1994; Werner \& \ Dreizler 1994)
and recently reported for one other hot DO (PG0108+101, Dreizler 2000)
but it is
always associated with the presence of iron. Furthermore, the measured Fe
abundance is typically larger than that of Ni, by factors between 1 and
20. Hence, it is very surprising, given the
detection of Ni in REJ0503$-$289, that we find no evidence at all of any
Fe. Not even coadding the regions of the strongest predicted Fe lines in
velocity space reveals the slightest hint of an absorption feature.
Indeed, this technique allows us to place a tighter upper limit on the
abundance of Fe, at $\approx 10^{-6}$, than imposed by the individual
lines. Thus, the abundance of Ni is greater than the abundance of Fe in
REJ0503$-$289, in the opposite sense to what is observed in the DA white
dwarfs, the Fe/Ni ratio being about one order of magnitude lower than found
in those stars. Interestingly, the only other DO star with
any iron group elements is PG1034+001.  KPD0005+5106 was
reported to have Fe VII features but GHRS and coadded IUE spectra do
not show these features (Werner \etal \  1996 and Sion \etal \  1997).
Dreizler \& \ Werner (1996) find log(Fe/He) $= -5$ and log(Ni/He) $ < -5$ in
PG1034+001.  For PG0108+101, Dreizler (2000) gives log(Fe/He) $= -4.3$
and log(Ni/He) $= -4.3$.
Thus, the only other DOs with detectable iron and nickel, 
have Fe/Ni ratios $> 1$, like those of the DA stars. 

The relative abundances observed in REJ0503$-$289 are clearly in
disagreement with the cosmic abundance ratio of Fe/Ni ($\approx
18$). Unfortunately, the predictions of radiative
levitation calculations do not offer much help in explaining the
observations. First, while Fe and Ni levitation has been studied in DA
white dwarfs, the predicted abundances are much larger than observed,
when all possible transitions (from the Kurucz line lists, Kurucz 1992)
are included in the calculations (Chayer
\etal \ 1994). Interestingly, better agreement is achieved, for Fe at least,
when the subset of lines available in TOPBASE (Cunto \etal \  1993)
is used and after several physical improvements to the calculations
(Chayer \etal \ 1995). Until recently, no similar calculations were
available for Ni. Although much of the radiative levitation work has
concentrated on DA white dwarfs, Chayer, Fontaine
\& \ Wesemael (1995) did deal with radiative levitation on He-rich
atmospheres, but only considering elements up to and including Fe. The
predicted Fe abundance for a star with the temperature and gravity of
REJ0503$-$289 is in excess of $10^{-4}$, two orders of magnitude above
the level observed by us. 

Recently, Dreizler (1999, 2000) has calculated NLTE model atmospheres taking
the radiative levitation and gravitational settling self-consistently into
account. In agreement with the results of Chayer, Fontaine
\& \ Wesemael (1995) the predicted iron abundance is far in excess of the
observed one. The observed Fe/Ni ratio, however, can be reproduced
qualitatively by these new models, which predict an excess of the nickel
abundance over the iron abundance by a factor of three, but
it is also clear that the stratified models do not work
very well for the DOs, which are best represented by chemically homogeneous
calculations.

This clear anomaly mirrors the comparison between predicted and observed
abundances for most of the heavy elements (see Table~\ref{mods}). 
Only the observed abundance of oxygen is close to its predicted
value. Consequently, it seems
reasonable to conclude that the theoretical calculations are deficient in
some way. The Chayer \etal \ (1995) 
make very clear statements about what physical
effects are considered by their work and what, for various reasonable
reasons, they do not deal with. Perhaps the most important limitation
is that the current published results for He-rich stars are for static
atmospheres whereas there is some evidence for active mass-loss
in He-rich objects, including REJ0503$-$289 (Barstow \& \ Sion 1994).
On the other hand, recent studies of DA white dwarf atmospheres 
show evidence of heavy element stratification
(Barstow \etal \ 1999; Holberg \etal \ 1999;
Dreizler \& \ Wolff 1999). This calls into question
the validity of trying to compare `abundances' determined from
homogeneous models with the radiative levitation predictions where
depth dependent elemental abundances are a direct result of the
calculations.

\section{Conclusion}

We have presented the first direct detection of nickel 
(Ni/He$=10^{-5}$) in the
photosphere of the hot DO white dwarf REJ0503$-$289
together with a new determination of $T_{\rm eff}$ and log g
utilising an objective spectral fitting technique. Nickel
has been seen previously in the atmospheres of hot H-rich white dwarfs,
but this is one of the first similar discoveries
in a He-rich object, detection of Ni in PG0108+101 having
also been recently reported by Dreizler (2000). It is also
one of a very small number of detections of Fe group elements in
any of the DO white dwarfs. A careful search for the presence of Fe
in the star only yields an upper limit of Fe/He$=10^{-6}$, implying
a Fe/Ni ratio a factor 10 lower than seen in the H-rich white dwarfs.
Although there are no published theoretical predictions, from radiative
levitation calculations, for the abundance of Ni in He-rich
photospheres the observed Fe abundance is some two orders
of magnitude below that expected. An explanation of the
observed heavy element abundances in this star clearly
requires new studies of the various competing effects that determine
photospheric abundance, including radiative levitation and
possible mass loss via winds. In addition, model atmosphere
calculations need to consider the possible effect of depth-dependent
heavy element abundances. Some work of this nature has already
been undertaken for DA white dwarfs but is only just beginning for He-rich objects. This work appears to be necessary to explain the continued problem
of the inconsistency between the results of the far-UV analyses and the
EUV spectrum, which cannot be matched by a model incorporating the abundances
measured here at the optically determined temperature, the predicted EUV flux
level exceeding that observed by a factor 2--3.
However, the early indication of studies using self-consistent stratified
models is that they do not match the observations very well.

\section*{Acknowledgements}

The work of MAB was supported by PPARC, UK, through an Advanced Fellowship. 
JBH and EMS wish to acknowledge support for this work
from NASA grant NAG5-3472 and through grant GO6628 from the
Space Telescope Science Institute, which is operated by the
Association of Universities for Research in Astronomy,
incorporated under NASA contract NAS5-26555. 
\hs \ data analysis in T\"ubingen is supported by the
DLR under grants 50 OR 96029 and 50 OR 97055.
Data analysis and interpretation were performed using 
NOAO \iraf , NASA HEASARC and Starlink software. We would like
to thank the support scientists at the Space Telescope Science
Institute for their help in producing  successul observations
of REJ0503$-$289.

\end{document}